\documentclass{JINST}

\title{UV Degradation of the Optical Properties of Acrylic for Neutrino and Dark Matter Experiments}

\author{B. Littlejohn$^a$\thanks{Corresponding
author.}~, K. M. Heeger$^a$\thanks{Corresponding
author.}~, T. Wise$^a$, E. Gettrust$^b$, and M. Lyman$^b$\\
\llap{$^a$}Physics Department, University of Wisconsin,\\
  Madison, WI, 53706, USA\\
\llap{$^b$}Madison West High School,\\
  Madison, WI, 53726, USA\\
  E-mail: \email{littlejohn@wisc.edu, heeger@wisc.edu}}

\abstract{UV-transmitting (UVT) acrylic is a commonly used light-propagating material in neutrino and dark matter detectors as it has low intrinsic radioactivity and exhibits low absorption in the detectors' light producing regions, from 350~nm to 500~nm.  Degradation of optical transmittance in this region lowers light yields in the detector, which can affect energy reconstruction, resolution, and experimental sensitivities.  We examine transmittance loss as a result of short- and long-term UV exposure for a variety of UVT acrylic samples from a number of acrylic manufacturers.  Significant degradation peaking at 343~nm was observed in some UVT acrylics with as little as three hours of direct sunlight, while others exhibited softer degradation peaking at 310~nm over many days of exposure to sunlight.  Based on their measured degradation results, safe time limits for indoor and outdoor UV exposure of UVT acrylic are formulated.}


\begin{document}

\section{Introduction}

Acrylic, or poly(methyl methacrylate), is a common material of choice for detectors and light guides in nuclear and particle physics experiments, as it is a relatively chemically inert, cheap, and strong material with low radioactivity and high optical clarity.  While laboratory-grade acrylics often transmit far into the UV spectrum \cite{Book}, commercial acrylics used in most large-scale scientific and non-scientific applications contain UV-absorbing additives to boost mechanical and optical stability under exposure to sunlight or UV irradiation \cite{AcrylicBook}.  Commercial acrylics not containing such additives are referred to as UV-transmitting (UVT) acrylics, and, while inferior to laboratory-grade PMMA, still retain a high degree of UV transmittance and can be manufactured on a large scale.  Hence, UVT acrylic is a natural choice for experiments seeking high transmission of UV and visible light over long distances.  In water Cherenkov detectors, such as the Sudbury Neutrino Observatory and the Super-Kamiokande outer detector, UVT acrylic can be used to contain water volumes and to propagate or wavelength-shift Cherenkov light, which increases in intensity with decreasing wavelength well into the UV range \cite{SNO, SuperK}.  Some liquid noble gas dark matter experiments, such as DEAP/CLEAN, will use UVT acrylic to pipe scintillation light shifted from XUV to near-UV wavelengths from a radiopure target volume to photomultiplier tubes \cite{DEAP, CLEAN, OrganicFluors}.

\begin{figure}[h]
\begin{center}
\includegraphics[width=110mm]{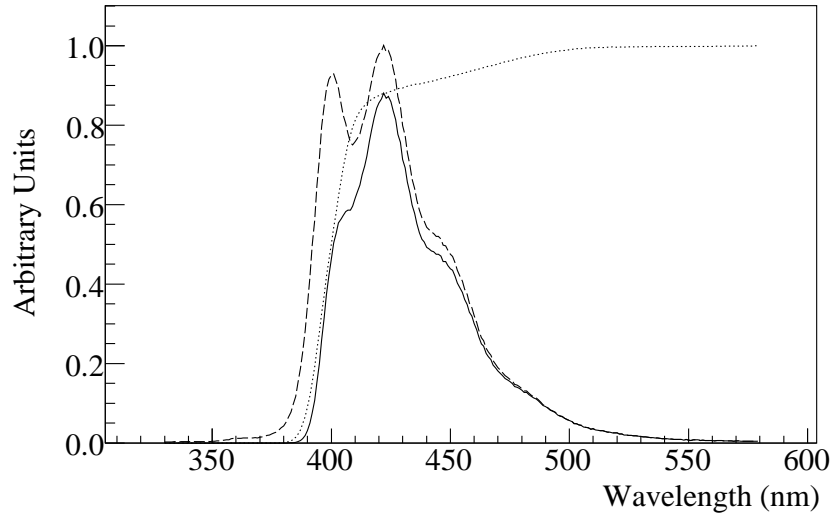}
\caption{A sample Daya Bay liquid scintillator emission spectrum (dashed); transmittance spectrum of 1~cm non-UVT acrylic (dotted); light yield based on these two values (solid).  The peak values of the two input spectra have been normalized to 1.  Sample emission spectrum information provided by \cite{Jun}.}
\label{fig:Emission}
\end{center}
\end{figure}

A wide array of organic liquid scintillator experiments also plan to use UV-transmitting acrylic.  The SNO+, Double Chooz, and Daya Bay experiments, as well as neutrino-based nuclear reactor monitoring systems, such as SONGS1, use organic liquid scintillators to convert energy depositions from inverse beta-decay reactions into scintillation light \cite{TDR, DoubleChooz, SNOPlus, ReactorMonitoring}.  Scintillators are chosen whose emission wavelengths match the wavelength ranges of the experiment's PMTs for maximum light collection.  A sample liquid scintillator emission spectrum from Daya Bay is shown in Figure \ref{fig:Emission}; the product of this spectrum and a non-UVT acrylic transmittance spectrum reveals about 21\% light loss in traversing 1~cm of acrylic.  If UVT acrylic is used, light loss due to acrylic traversal will be less than a few percent.

Previous studies indicate that exposure of laboratory-grade PMMA to high amounts of specific wavelengths of UV radiation result in photodegradation of the polymer, including main-chain scission and detachment of methyl and esther side-groups \cite{PMMADeg}.  The subsequent introduction of these new products in the material results in increased UV absorption, with a band peaking at 285~nm \cite{OldPMMATrans, NewPMMATrans}.  It has been demonstrated that long-term UV exposures via solar irradiation have also resulted in optical degradation of some commercial acrylics \cite{UVExposure}.  In addition to lowering the total light yield for all experiments, this material behavior is problematic for high-precision experiments that rely on well-known, uniform light yields to minimize detector-related systematic uncertainties.  If a detector's acrylic optical properties degrade after final optical measurements have been made in the laboratory, this would lead to a discrepancy between a detector's actual optical properties and its optical model.  This would result in an incorrect modeling of the detector's light yield, causing improper energy reconstruction.

This study's goal is to establish relationships between UV exposure and degradation of acrylic transmittance for various commercial acrylics.  Previous studies have done this for exposures with durations of several years.  However, all aforementioned particle and nuclear physics experiments operate in the absence of external optical or UV light and will only be exposed to damaging UV radiation during construction, transportation and any above-ground storage they may experience.  Thus, it is necessary to examine any effects of high-intensity UV irradiation on commercial acrylics for periods on the order of weeks or months.  Degradation of optical properties in the 360-500~nm range are of particular concern, as this is the region of highest light emission and PMT quantum efficiency for most of the experiments.  The results presented here are obtained from studies of candidate acrylic materials for the Daya Bay reactor antineutrino experiment.

\section{Experimental set-up}

Tested acrylic samples were obtained from two US commercial manufacturers, Reynolds Polymer Technology \cite{ReynoldsWeb} and Spartech Polycast \cite{PolycastWeb}, and one foreign manufacturer, YenNan Taiwan \cite{YenNan}.  Reynolds Polymer Technology provided a number of slush-cast samples made at their Grand Junction, Colorado facility that differed only in the polymer used in the starting slush.  The polymers used were Lucite, ACP-10, and another unknown polymer \cite{Reynolds}.  Another sample was manufactured by Reynolds Thailand using a pure monomer cast.  The Daya Bay experiment's outer acrylic vessels will be constructed from a combination of Polycast and Reynolds acrylic.  The YenNan sample material was used to construct a prototype of Daya Bay's inner acrylic vessels.  Once received, samples were cut to smaller sizes and polished to 0.3~$\mu$m grain at University of Wisconsin.  Sample details are listed in Table ~\ref{tab:samples}.

\begin{table}[h]
\begin{center}
\begin{tabular}{|c|c|c|c|c|c|}
\hline
Manufacturer & Sample & Pathlength & Short-Term & Long-Term & Equivalent UV \\
Name & Name & (mm) & Exposure & Exposure & Dosage (J/cm$^{2}$) \\
 & & & Time (hr) & Time (d) & \\
\hline
Polycast & \#18 & 9.5 & 19.5 & - & 260 \\
 & \#7 & 9.5 & - & 25 & 2192 \\
\hline
Reynolds &  \#17 & 9.5 & 24 & 5 & 760 \\
 &  \#1 & 9.5 & 10.5 & 5 & 598 \\
Reynolds & ACP-10 \#26 & 10 & 4 & 14 & 980 \\
Reynolds & Lucite \#25 & 10 & 4 & - & 14.5 \\
Reynolds & Thailand \#24 & 10 & 4 & 14 & 980 \\
\hline
YenNan & Taiwan \#19 & 10 & 12 & - & 180 \\
 & Taiwan \#20 & 10 & - & 22 & 1660 \\


\hline
\end{tabular}
\caption{UV exposure and sample information.  Pathlength indicates the length of acrylic sample light must traverse during a transmittance measurement.  Note that some Reynolds samples were subjected to short-term testing and then long-term testing directly afterward; for these samples, the given UV dosage is the total accumulation over all testing.}

\label{tab:samples}
\end{center}
\end{table}

\begin{figure}[h]
\begin{center}
\includegraphics[width=150mm]{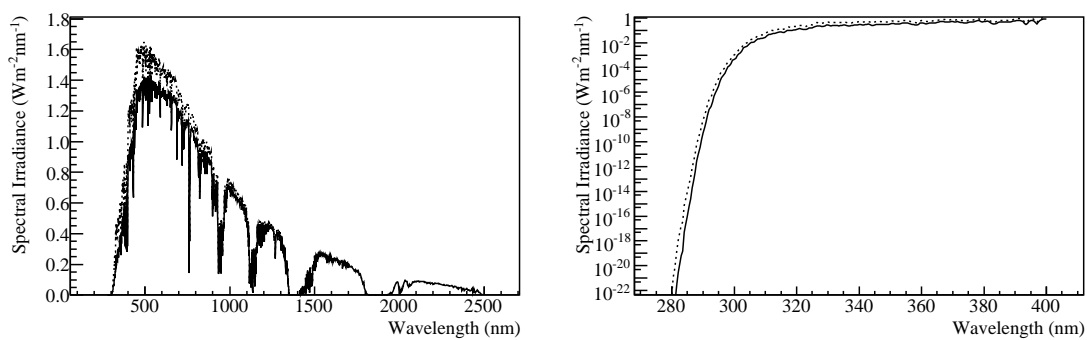}
\caption{Left: total solar UV/VIS/NIR spectrum  according to ASTM standard G173-03 \cite{ASTM}.  Right: total solar UV spectrum.  Spectra are given for hemispherical terrestrial measurements (dotted), and direct solar plus circumsolar measurements (solid).}
\label{fig:sun}
\end{center}
\end{figure}

UV exposure was achieved via outdoor solar irradiance.  The total solar emission spectrum can be seen in Figure \ref{fig:sun}, along with solar UV emission.  During exposure, samples were mounted on a flat, untilted, non-reflective surface and then exposed to the sun in Madison, Wisconsin for either short-term or long-term testing.  Short-term exposures were 3-4 hours in length and were repeated for up to 24 hours.  Long-term exposures ranged from 5-13 days in length.  Tests were performed in the period between August 14 and September 15, 2008.  Short-term exposures were achieved during sunny or mostly sunny weather conditions.  Long-term exposures included periods of rain, clouds, sun, and night.

Two UV filters with different transmittance cut-off frequencies were used during some UV exposures: a Roscolux \#3314 UV Filter had a cutoff wavelength of about 400~nm, while a 3~mm glass pane served as a UV filter with a 320~nm cutoff wavelength.  By exposing quickly-degrading samples to varying degrees of UV-filtered sunlight, non-damaging and damaging radiation wavelengths could be identified.  This test was performed on Reynolds \#1 sample only.

Total UV (UVA + UVB) solar irradiance was measured during sunny and cloudy periods at the sample exposure location with a Solartech Solarmeter Model 5.0 UV Meter \cite{UVMeter}.  Mid-day UV irradiance values reached a maximum of 5.5 mW/cm$^{2}$.  Cloud or shade cover yields irradiance values of 0.5 - 1.5 mW/cm$^{2}$, while fully cloudy or rainy conditions yield values of less than 0.1 mW/cm$^{2}$.  As a reference, ASTM standard sun values give UV exposures (280~nm-400~nm) of 9.1 mW/cm$^{2}$ for hemispherical acceptance at $37^{\circ}$ tilt and 6.1 mW/cm$^{2}$ for direct and circumsolar angular acceptance \cite{ASTM}.

Solar UV radiation levels vary greatly with weather and time of day, making long-term UV exposure values harder to quantify.  UV meter readings during cloudy, sunny, mid-day, and early morning periods correlate well with real-time solar radiation data provided by the University of Wisconsin-Madison Rooftop Instrument Group (RIG) \cite{Weather}. While the UV meter readings must be made manually, RIG equipment automatically provides two solar radiation readings per minute, twenty-four hours a day.  Thus, total UV exposure values for each sample were obtained by normalizing RIG solar radiation data to UV meter readings.  The average UV dosage during a three-hour exposure is $\sim$40~J/cm$^{2}$.  Some systematic uncertainty in total UV exposure is introduced because the RIG data is not available in raw format and is taken from the published graphical format.

Transmittance spectra of unexposed and exposed samples were taken using an SI Photonics Model 440 UV/Vis spectrophotometer \cite{Spectrometer}.  The spectrometer uses a fiber-optic light delivery and collection system, as well as a CCD detector, which allows for samples of large or varied sizes to be measured outside of a dark-box environment. Non-parallel sample surfaces and varying sample thickness result in imperfect light refocusing, resulting in varying degrees of light collection in the fiber-optic system.  Lack of consistent light collection is the main contributor to the systematic error in the transmittance measurements.  Another significant source of uncertainty is uneven sample surface quality; despite polishing to 0.3~$\mu$m grain, light surface scratching was still visible on most samples.  In addition, our polishing technique evolved over the course of the experiment, likely causing slight differences in surface quality from sample to sample.  Transmittance values at 360, 380, 400, and 500~nm are mainly presented in this report, as they are wavelengths that span the typical emission spectrum of liquid scintillators and are of interest in the Daya Bay experiment.

\section{Results}
\subsection{Degradation under short-term UV exposure}

\begin{figure}[h]
\begin{center}
\includegraphics[width=150mm]{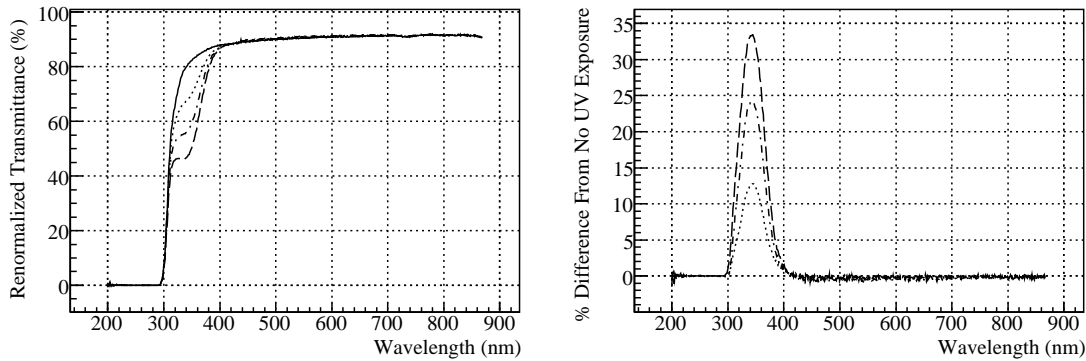}
\caption{Left: transmittance spectrum for Reynolds sample \#17.  Right: Difference in transmittance of sample before and after UV exposure.  Solid - No Exposure; Dotted - 32.9~J/cm$^{2}$; Dot-Dashed - 66.8~J/cm$^{2}$; Dashed - 115~J/cm$^{2}$.}
\label{fig:17UV}
\end{center}
\end{figure}

Short-term tests consist of samples undergoing single or repeated exposures of 4 hours or less.  The results of short-term exposure studies are summarized in Table \ref{tab:shortterm}.  The Reynolds \#17 sample was submitted to 7 exposures of 3-4 hours and received the highest UV dosage of any sample in the short-term test.  After exposure, the sample showed no signs of coloration, crazing, or surface degradation.  The transmittance spectrum of the Reynolds \#17 acrylic sample for selected UV dosages can be seen in Figure \ref{fig:17UV}.  In this sample, significant degradation of specular transmittance can be seen in the near-UV range for the lowest dosage level, 32.9 J/cm$^{2}$.  The difference between exposed and unexposed transmittances, hereafter referred to as transmittance degradation, as plotted in Figure \ref{fig:17UV}, clearly exhibits a single peak of increasing amplitude.  A gaussian fit gives this peak a mean value of 343~nm and a half-width of 20.3~nm.  The mean value remains consistent with increasing dosage while $\sigma$ increases on the order of a few nm.

\begin{table}[h]
\begin{center}
\begin{tabular}{|c|c|c|c|}
\hline
 Sample & \# of Exposures & Total Dosage (J/cm$^{2}$) & $\Delta$T$_{360nm} (\%)$ \\
\hline
Reynolds \#17 & 7  & 296 & -47.5 \\
Polycast \#18 & 6 & 263 & +1.8 \\
YenNan Taiwan \#19 & 4 & 181 & -3.2 \\
Reynolds Thailand \#24 & 1 & 63 &  +3.8 \\
Reynolds Lucite \#25 & 1 & 63 & -15.5 \\
Reynolds ACP-10 \#26 & 1 & 63 & -1.8 \\
\hline
\end{tabular}
\caption{Short-term UV exposure results for Reynolds, Polycast and YenNan samples.  $\Delta$T$_{360nm}$ refers to the change in transmittance from non-exposure at 360~nm after at the listed UV dosage.  Fluctuations of less than 4\% are unresolvable due to systematic uncertainties from the spectrometer and variation in sample surface preparation.}
\label{tab:shortterm}
\end{center}
\end{table}

\begin{figure}[h]
\begin{center}
\includegraphics[width=90mm]{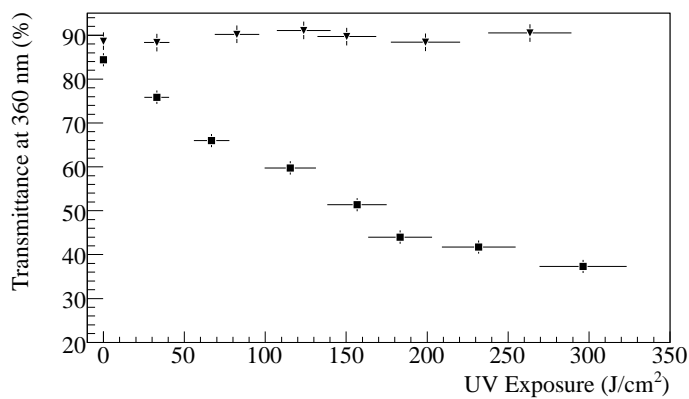}
\caption{Transmittance at 360~nm plotted versus UV dosage.  Triangles - Polycast \#18; Squares - Reynolds \#17.  Vertical error bars indicate transmittance measurement uncertainty from sample surface imperfections and imperfect light collection, while the horizontal error bars show the uncertainty in UV exposure resulting from varying weather conditions.}
\label{fig:shorttermfig}
\end{center}
\end{figure}

The degradation at 360~nm as a function of UV dosage is shown in Figure \ref{fig:shorttermfig}.  As suggested by Figure \ref{fig:17UV}, transmittance at 360~nm continues to degrade with increasing UV dosage, although the pace of degradation slows.  The data fits an exponential decay with an exposure constant on the order of 220~J/cm$^{2}$, with a horizontal asymptote at 20\% transmittance.  The uncertainty in exposure is the result of a 1 mW/cm$^{2}$ uncertainty in maximum solar luminosity, a 0.6 mW/cm$^{2}$ uncertainty during shaded exposure, both very conservative values, and a 5-minute period of uncertainty in luminosity for every 30 minutes of full sunlight to account for periods of brief cloud cover.  To allow for relative comparisons despite uneven spectrometer light collection, each individual sample's spectra were normalized to match transmittances over the long period of high transmittance in the visible spectrum.  The uncertainty in the transmittance for a sample was defined as the deviation of this normalization constant from unity.

The Reynolds Lucite \#25 sample, also tested in this manner, albeit with fewer exposures, exhibited similar behavior.  As with Reynolds \#17, degradation occurred at 343~nm with slightly wider half-width of 20.7~nm.  The decay rate at this wavelength is also comparable: at a UV dosage of 66.8~J/cm$^{2}$ and 63~J/cm$^{2}$ for Reynolds \#17 and Reynolds Lucite, respectively, Lucite showed 22.5\% degradation at 343~nm while Reynolds showed 24.5\%.  Meanwhile, the two other Reynolds samples were stable, showing no appreciable short-term degradation.

The Reynolds \#1 sample, from the same cast acrylic block as \#17, showed similar degradation under one short-term exposure, as expected.  Three subsequent short-term exposures and one 5-day long-term exposure were then performed under 3~mm glass UV filtration.  No degradation was observed during any of these filtered tests.  Thus, it appears that UV radiation less than 320~nm in length is responsible for short-term degradation of acrylic, in agreement with previous literature on UV degradation of PMMA.

All other samples showed no resolvable signs of short-term degradation.  It appeared that some samples may have been experiencing a slight decay over a broad range of near-UV wavelengths, but any change was smaller than the systematic uncertainty of the experimental setup.  The clear degradation of Reynolds \#17 is compared to the stability of transmittances for Polycast \#18 at 360~nm in Figure ~\ref{fig:shorttermfig}. All non-degrading samples also showed no visible signs of yellowing, crazing, or surface degradation during all short-term testing.

\subsection{Long-term degradation}

\begin{figure}[h]
\begin{center}
\includegraphics[width=150mm]{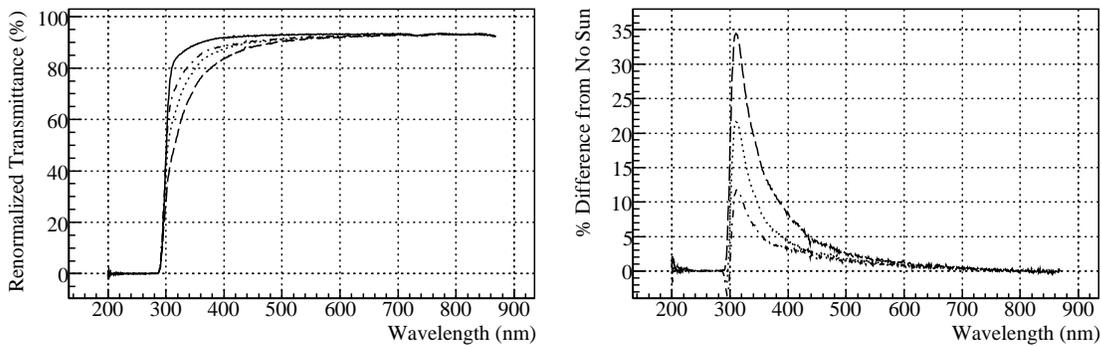}
\caption{Right: transmittance spectrum for Polycast sample \#7.  Left: Difference in transmittance of sample before and after UV exposure.  Solid - No Exposure; Dotted - 752~J/cm$^{2}$; Dot-Dashed - 1211~J/cm$^{2}$; Dashed - 2192~J/cm$^{2}$.}
\label{fig:7UV}
\end{center}
\end{figure}

The effect of long-term exposures of up to 24 days of solar irradiation on acrylic was also studied.  Long-term exposure results are summarized in Table ~\ref{tab:longterm}.  The Polycast \#7 sample received the greatest long-term UV dosage; its transmittance spectra for increasing dosage levels is shown in Figure ~\ref{fig:7UV}.  The change in transmittance over the spectrum is not gaussian, but shows instead a sharp peak at 310~nm, followed by a long tail extending out past 450~nm.  Change in transmittance remains above 5\% until about 435~nm for maximum UV dosage.  The YenNan \#20 sample's transmittance change is similar, with a broader peak at around 305~nm.  Neither sample showed any visible signs of yellowing or degradation.

\begin{table}[h]
\begin{center}
\begin{tabular}{|c|c|c|c|}
\hline
 Sample & Exposure Time (d) & Total Dosage (J/cm$^{2}$) & $\Delta$T$_{360nm} (\%)$  \\
\hline
Polycast \#7 & 25 & 2192 & -13.8 \\
Reynolds \#17 & >5 & >755 & -62.2 \\
YenNan \#20 & 19 & 1667 & -8.9 \\
Reynolds Thailand \#24 & 14 & 981 & -1.1 \\
Reynolds ACP-10 \#26 & 14 & 981 & -16.3 \\
\hline
\end{tabular}
\caption{Long-term UV exposure results for Reynolds, Polycast and YenNan samples.  Fluctuations of less than 3-4\% cannot be resolved due to systematic uncertainties from the spectrometer and variation in sample surface preparation.}
\label{tab:longterm}
\end{center}
\end{table}

\begin{figure}[h]
\begin{center}
\includegraphics[width=150mm]{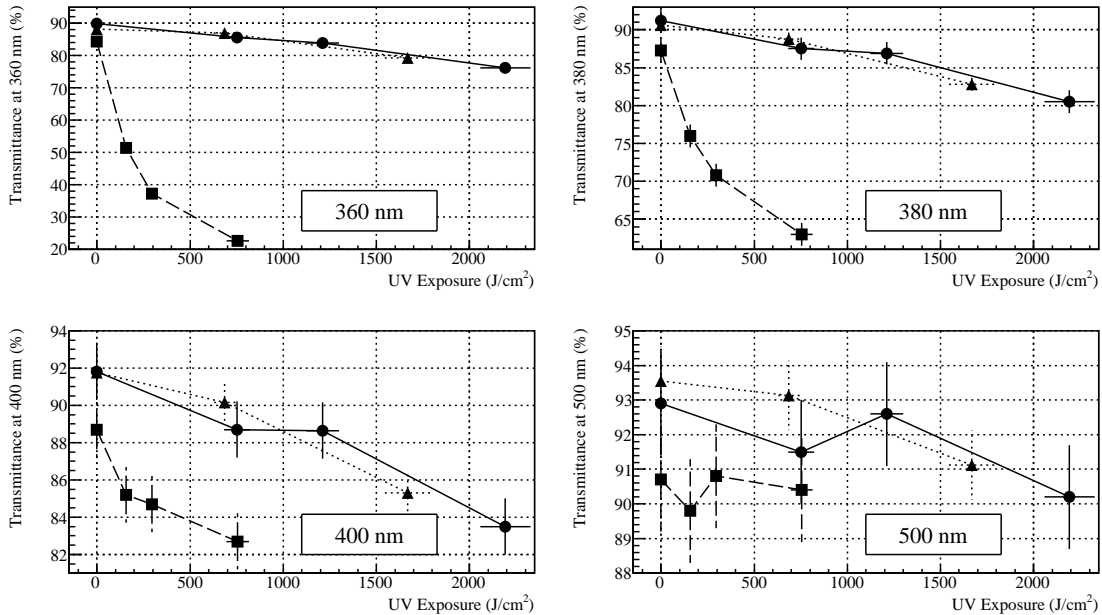}
\caption{Transmittance plotted versus UV dosage for four wavelengths.  Circles - YenNan \#20; Triangles - Polycast \#7; Squares - Reynolds \#17. Vertical error bars indicate transmittance measurement uncertainty from sample surface imperfections and imperfect light collection, while the horizontal error bars show the uncertainty in UV exposure resulting from varying weather conditions.}
\label{fig:LongTermExposure}
\end{center}
\end{figure}

Rate of long-term degradation of Polycast and YenNan samples can be examined by plotting transmittance at 360~nm versus total UV dosage, as was done with short-term exposure data.  In Figure ~\ref{fig:LongTermExposure}, one can see the difference in degradation rate between Reynolds \#17 and the Polycast and YenNan samples.  A linear fit to the Polycast and YenNan samples yields a degradation rate of 6.2~cm$^{2}$/kJ and  5.5~cm$^{2}$/kJ, respectively.  The units on this rate are percent loss in transmittance over exposure in kJ/cm$^{2}$.  At this rate, a Polycast sample would need around 50 times more UV exposure than a Reynolds \#17 sample to degrade transmittance to 90\% of its initial value at 360~nm.

This rate decreases substantially for higher wavelengths' transmittance.  At 380~nm and 400~nm, degradation rates for Polycast are 4.7~cm$^{2}$/kJ and 3.6~cm$^{2}$/kJ, respectively.  At 500~nm, any change in transmittance for all samples is within the statistical uncertainty of the experimental setup.  Transmittance uncertainty was calculated in a similar manner as the short-term tests.  Uncertainty in exposure is dominated in long-term tests by the conservative uncertainty in the maximum solar luminosity.

The Reynolds ACP-10 \#26 sample experienced similar long-term degradation, with change in transmittance peaking at 311~nm.  The tail is longer in the ACP-10 sample, however.  In Polycast, at a dosage of 1211~J/cm$^2$, the tail dips below 5\% at 400~nm, while in ACP, a smaller dosage, 981~J/cm$^2$ yields a tail dipping below 5\% at 450~nm.  Also in contrast to samples \#7 and \#20, rate of degradation is higher in the Reynolds ACP-10 sample.  At 360~nm, Reynolds ACP-10 degraded 16.3\% with 981~J/cm$^2$ of UV exposure, giving a degradation rate of about 16.7~cm$^2$/kJ, about 2.7 times greater than that of Polycast.  Still, the short-term UV stability of the ACP-10 samples makes them suitable for use in the Daya Bay experiment.  Slight yellowing of the ACP-10 sample was visible after maximum UV exposure.

The Reynolds Lucite sample did not display any change in transmittance greater than the systematic uncertainty in the transmittance measurement.  It seems likely, however, that given a greater exposure, the sample would undergo long-term degradation similar to that of the Polycast, YenNan, and Reynolds ACP-10 acrylics.  It is not possible to distinguish similar behavior in the Reynolds \#17 sample, as the transmittance loss described when discussing short-term degradation washes out degradations of any other kind that would take place over a longer period of time.

As UV filtration was only done with Reynolds Acrylic \#1, a short-term degrading sample, the wavelengths responsible for the long-term degradation pattern seen in Polycast, YenNan, and Reynolds ACP-10 samples are not known.

\section{Discussion}

Two separate decay processes are being observed in the different commercial acrylic samples, one fast decay causing gaussian-shaped changes in transmittance peaked at 343~nm, and one slower decay causing changes in transmittance that peak at 310~nm and have a tail extending into the visible range. 

\subsection{Observations}

The two observed decay processes are caused by photochemical degradation of different molecules or of different regions of the same molecule in UVT acrylic.  Previous polymer science literature cites main-chain scission of the PMMA molecule, as well as detachment of its methyl and esther side groups as the cause of degradation of laboratory acrylic optical properties.  It is unclear from this study or from previous polymer science literature which of these chemical processes, if any, are responsible for the optical decay processes observed in this study.  It is possible that the degradations of industrial acrylic could be caused by chemical additives not found in laboratory-grade acrylics.  The polymers used in Reynolds \#17 and Reynolds Lucite acrylics could contain additives not present in the polymers of other samples in this or previous polymer science studies whose degradation causes their unique short-term decay mode.  It is also possible that the process used to make these polymers causes their constituent PMMA chains to be more susceptible to photolysis at certain places along the molecular chain.  Not enough is known about commercial acrylic casting processes or industrial polymer synthesis to make a conclusive statement; much of the information about industrial acrylic production processes are proprietary and not available to the public.

It is interesting to note that all transmittance spectra differ greatly from previous polymer science and chemistry literature regarding PMMA, even before UV exposure.  All samples tested exhibited below 1\% transmittance below 270~nm, while in previous studies \cite{OldPMMATrans, NewPMMATrans}, lab-synthesized PMMA exhibits transmittances well above 25\% below 270~nm.  In previous studies, UV degradation is greatest in the far UV region, from 230-290~nm.  It is possible that commercial-grade UVT acrylics contain additives to absorb these wavelengths before they are absorbed by PMMA and degrade its chemical structure, and subsequently its optical and structural properties.  If this is the case, the long-term degradation exhibited in the tested samples, which occurs at the very beginning of the transmitting region of the sample, may be the outlying effects of UV degradation described in previous literature.  In contrast, the short-term degradation experienced by Reynolds \#17 and Lucite, which peaks in the 343~nm range, is not described in any previous studies of pure laboratory PMMA.

\subsection{Acceptable exposure times}

Reynolds \#17 and Reynolds Lucite acrylic samples display the gaussian-shaped short-term degradation.  If high, well-known UV transmittance values are desired, objects made of these acrylics kept in outdoor areas for periods longer than an hour may experience significant degradation.  Indoor construction or long-term storage facilities with extremely low UV exposure will further reduce the possibility of degradation.  In the Reynolds Polymer Technology (RPT) construction facility in Grand Junction, Colorado, which is likely similar in lighting to most normal factory facilities, indoor UV levels in areas free of direct sunlight are in the neighborhood of 4-12 $\mu$W/cm$^{2}$.  If we assume 5\% transmittance loss at 360~nm to be the maximum acceptable degradation, this corresponds to a factory exposure time of about 16 days.  Polycast and YenNan acrylic samples exhibited the lowest transmittance loss, with a rate of about 6.2~cm$^{2}$/kJ and 5.5~cm$^{2}$/kJ.  Experimental parts constructed with this material could be kept outdoors for many days without serious degradation, although it is not recommended.  If a 5\% maximum transmittance loss limit is again applied, indoor construction and storage at the RPT facility would be safe for as long as 800 days.  Reynolds ACP-10 lasts approximately 290 days under similar lighting.   These values are estimates, and may be lower or higher based on the emission spectra of the work lights in construction and storage facilities.

\subsection{Protection methods against the UV degradation of acrylic material}

Some remedies can be applied if further construction or storage time under these lighting conditions is needed.  UV filters can be used to block the UV component of factory lights to an acceptable level.  Tarped or similarly covered storage reduces UV exposure over long time periods to negligible levels.  Intense UV sources, greater than 100 mW/cm$^2$ at close-range, are also sometimes used to cure and strengthen acrylic structures.  It is recommended that such lights not be used on or in the vicinity of UVT acrylic.  For special applications such as the Daya Bay acrylic target vessels Reynolds Polymer Technology has developed a bonding process that does not reuly on the curing of bonds with UV light sources.  In addition, exposure of UVT acrylic to direct sunlight for long periods of time in a factory setting will have similar detrimental effects to those described in this paper.  Installing doors or walls in fabrication, construction, and storage facilities can minimize exposure to direct sunlight.

\section{Conclusion}

UV-transmitting acrylic is used in a number of neutrino and dark matter experiments as a light-transmitting container for the detector's target material and as a light guide. The low radioactivity of acrylic and its formable and machineable  properties make it a material of choice for many experiments including the Daya Bay reactor antineutrino experiment.  Using UV-Vis spectrometry, the optical stability of UVT acrylic samples exposed to direct solar radiation was determined for exposure periods from 3 hours to 24 days.  We have observed short-term and long-term degradation effects at $\sim$343~nm for short-term exposure and $\sim$310~nm for long-term exposure.  Tails from these degradation peaks extend into the near-UV and visible region, which could negatively affect energy reconstruction and resolution in experiments using the degraded acrylic.  Because of its high transmittance and low optical degradation, Polycast, Reynolds ACP-10, and YenNan acrylics are acceptable for use in Daya Bay and other experiments.  The Daya Bay experiment will use Polycast acrylic sheets to construct its 18~mm outer acrylic vessel walls, and cast Reynolds ACP-10 acrylic for the thicker acrylic vessel endcaps and support structures.  Tests similar to the one described in this paper can be performed on other acrylics to determine their suitability for use in neutrino and dark matter experiments.

UV filtration of factory lighting, tarped or covered storage, and construction of doors or walls to block direct solar exposure in a factory setting are all effective precautions to protect UVT acrylic from UV exposure during construction and storage of detector parts.  Outdoor exposure or UV curing of any kind is to be avoided, if possible.  If not possible, any UV curing regimen or period of outdoor exposure is to be minimal and closely planned.

\acknowledgments

This work was done under DOE contract and with support of the University of Wisconsin Foundation as part of R\&D for the Daya Bay reactor antineutrino experiment. We are grateful to Reynolds Polymer Technology, Inc., of Grand Junction Colorado for their producing and providing a variety of acrylic samples, both UVT and non-UVT for this study, and for their collaboration in developing the acrylic target vessels for the Daya Bay reactor antineutrino experiment.  We also thank Jun Cao from the Daya Bay collaboration at the Institute of High Energy Physics, Beijing, China for providing us with a sample Daya Bay liquid scintillator emission spectrum.  We acknowledge support of the QuarkNet program that provided E.G. and M.L with the opportunity to participate in the early stages of this work.

\end{document}